\documentclass[twocolumn,showpacs,preprintnumbers,amsmath,amssymb]{revtex4}
\usepackage{graphicx}
\usepackage{bm}

\def\itmb{\begin{itemize}}
\def\itme{\end{itemize}}
\def\enmb{\begin{enumerate}}
\def\enme{\end{enumerate}}
\def\eqnb{\begin{equation}}
\def\eqne{\end{equation}}
\def\PTP{Prog. Theor. Phys.(Kyoto)}

\def\NPB{{Nucl. Phys.} {\bf B}}
\def\PLB{{Phys. Lett.} B}
\def\PRL{Phys. Rev. Lett.}
\def\PRD{{Phys. Rev.} D}
\def\PRB{{Phys. Rev.} B}

\begin{document}
\newcommand{\ttbs}{\char'134}

\title{Effects of the quark field on the ghost propagator 
of Lattice Landau Gauge QCD}
\author{Sadataka Furui}
\email{furui@umb.teikyo-u.ac.jp}
\homepage{http://albert.umb.teikyo-u.ac.jp/furui_lab/furuipbs.htm}
\affiliation{%
School of Science and Engineering, Teikyo University, Utsunomiya 320-8551 Japan.
}%
\author{Hideo Nakajima}
\email{nakajima@is.utsunomiya-u.ac.jp}
 
\affiliation{
Department of Information Science, Utsunomiya University, Utsunomiya 321-8585 Japan. 
}%

\date{\today}

\begin{abstract}
Infrared features of the ghost propagator of color diagonal and color antisymmetric ghost propagator of quenched SU(2) and quenched SU(3) are compared with those of unquenched Kogut-Susskind fermion SU(3) lattice Landau gauge.  

We compare 1) the fluctuation of the ghost propagator, 2) the ghost condensate parameter $v$ of the local composite operator (LCO) approach and 3) the Binder cumulant of color anti-symmetric ghost propagator between quenched and unquenched  configurations.

The color diagonal SU(3) ghost dressing function of unquenched configurations has weaker singularity  than the quenched configurations. In both cases fluctuations become large in $q<0.5$GeV. 

The ghost condensate parameter $v$ in the ghost propagator of the unquenched MILC$_c$ configuration samples is $0.002\sim 0.04$GeV$^2$ while that of the SU(2) PT samples is consistent with 0.  

The Binder cumulant defined as $U(q)=1-\frac{1}{3}\frac{< \vec\phi^4 >}{(< \vec\phi^2 >)^2}$ where $\vec \phi(q)$ is the color anti-symmetric ghost propagator measured by the sample average of gauge fixed configurations via parallel tempering method becomes $\sim 4/9$ in all the momentum region. 
 The Binder cumulant of the color antisymmetric ghost propagator of quenched SU(2) can be explained by the 3-d gaussian distribution, but that of the unquenched MILC$_c$ deviates slightly from that of the 8-dimensional gaussian distribution. 

The stronger singularity and large fluctuation in the quenched configuration could be the cause of the deviation of the Kugo-Ojima confinement parameter $c$ from 1, and the presence of ordering in the ghost propagator of unquenched configurations makes it closer to 1.
\end{abstract}

\maketitle
\section{Introduction}
Infrared features of the ghost propagator are important in the analysis of color confinement mechanism and the running coupling. Kugo and Ojima\cite{KO} considered the two point function connected by the ghost propagator and expressed  the confinement criterion as
\begin{equation}
1+u(0)=1-c=\frac{Z_1}{Z_3}=\frac{\tilde Z_1}{\tilde Z_3}=0
\end{equation}
at the renormalization point $\mu=0$\cite{kugo}.
Here $Z_1$ and $\tilde Z_1$ are the vertex renormalization factor of the triple gluon vertex and the ghost anti-ghost gluon vertex, respectively and $Z_3$ and $\tilde Z_3$ are the wave function renormalization factor of the gluon and the ghost, respectively. 

If $\tilde Z_1$ is finite, divergence of $\tilde Z_3$ is a sufficient condition of the color confinement.   The lattice data suggest that $\tilde Z_3$ is infrared divergent, but its singularity is
not strong enough to hinder the running coupling measured as
\begin{equation}\label{alpha}
\alpha_s(q)=\frac{g_0^2}{4\pi}\frac{Z(q^2){ G(q^2)}^2}{{\tilde Z_1}^2}\sim \alpha_s(\Lambda_{UV}) q^{-2(\alpha_D+2\alpha_G)},
\end{equation}
approach zero in the infrared\cite{unquench}. Here $Z(q^2)$ and $G(q^2)$ are the gluon dressing function and the ghost dressing function, respectively. The same observation is reported in\cite{ilgen}.

The ghost propagator in the infrared region was investigated by several authors.  Common findings are that it is more singular than $q^{-2}$ and that in the infrared region its statistical fluctuation is large probably due to presence of Gribov copies\cite{FN04,FN05,muell,bclm,Orsay}.  In the quenched $32^4, 48^4$ and $56^4$ SU(3) lattice simulation, the color diagonal ghost propagator showed singularity of $q^{-2-\alpha_G}$ with $\alpha_G\sim 0.25$. In the Dyson-Schwinger (DS) approach, the infrared power
behavior of the ghost propagator and the gluon propagator $q^{-2-\alpha_D}$ have the relation $2\alpha_G+\alpha_D=0$ and the lattice data are consistent with this ansatz in $q>1$GeV region. As the magnitude of $\alpha_D$, Dyson-Schwinger approach\cite{vS} and  Langevin approach\cite{Zw} predicts -0.59, while the lattice data and DS approaches\cite{Blo,Kond} predicts -0.5.  If $\alpha_D$ is smaller than -0.5 the gluon propagator in the infrared vanishes and the Gribov-Zwanziger's conjecture on the color confinement of the gluon becomes satisfied. Recent detailed analysis of the finite size effect in the lattice confirms that infrared limit of $-\alpha_D$ in the DS approach $\kappa=0.5$ is compatible with the lattice data\cite{Orsay3,OS}. The relation $2\alpha_G+\alpha_D=0$ suggests presence of an infrared fixed point\cite{vS}. The infrared finite quark wavefunction renormalization $Z_\psi$ of unquenched simulation\cite{FN06} also suggests that the running coupling is not infrared vanishing.

We cannot measure the ghost propagator at zero momentum, since we evaluate it with the condition that it is zero-mode-less.
Thus the infrared power fitted at finite lattice momentum $\alpha_G$ cannot predict the power behavior of the ghost propagator near momentum 0 i.e. the index $\kappa$. 

In \cite{unquench}, we observed that the Kugo-Ojima confinement criterion is satisfied in the unquenched simulation but not in the quenched simulation of lattice sizes up to $56^4$. In order to study the role of fermion in the color confinement, we consider the BRST(Becchi-Rouet-Stora-Tyutin) quartet mechanism\cite{KO, adfm}.
 
In the BRST formulation\cite{KO}, unphysical degrees of freedom are confined by the quartet mechanism.  In the pure QCD in the Landau gauge, one can construct BRST quartet as

\[
\begin{array}{ccccccc}
A_\mu & \to & D_\mu(A) c & \to & 0 &  & \\
        &  &A_\mu \bar c &\to  & D_\mu(A)c \bar c - A_\mu B & \to &0 . 
\end{array}
\]
Here the arrow implies the BRST transformation $\delta_B$ and $B$ is the Nakanishi-Lautrup auxiliary field.
The transverse gluon state $A_\mu$ is a BRST parent state of a daughter state $D_\mu(A)c$ and the state with opposite ghost number of the $D_\mu(A)c$ i.e. $A \bar c$ becomes a parent state, whose daughter and the above three states construct a quartet. 
 
Inclusion of the fermion field $\psi$ allows to construct another BRST quartet as
\[
\begin{array}{ccccccc}
\psi & \to & -\psi c & \to & 0 & & \\
      &     & \psi \bar c & \to & -\psi c \bar c-\psi B & \to & 0.
\end{array}
\]
The Dirac fermion state $\psi$ is a BRST parent state of $\psi c$ and the state with opposite ghost number state of $\psi c$ is $\psi \bar c$, which becomes a parent state of the BRST partner that construct a quartet.

Inclusion of fermion gives more restiction on the degrees of freedom of the ghost and it may change the fluctuation of the ghost propagator.

Another current problem concerning the ghost propagator is the possibility of the ghost condensates.
In the lattice Landau gauge QCD simulation, presence of $A^2$ condensates was suggested\cite{Orsay,FN04,FN05,unquench,lat05_a}. Since $A^2$ is not BRST invariant, a mixed condensate i.e. a combination with ghost condensates
\begin{equation}
 \int  \langle tr_{G/H}[\frac{1}{2}{\mathcal A}_\mu{\mathcal A}^\mu-\xi i{{\mathcal C}}\bar{\mathcal C}]\rangle d^4 x
\end{equation}
was proposed\cite{kondo,grip} as on-shell BRST invariant, i.e. invariant for the $B$ field that satisfies
\begin{equation}
B^a=-\frac{1}{\xi}\partial_\mu A^{a \mu}+i\frac{g}{2}f^{abc}c^b\bar c^c.
\end{equation}
 Here $G/H$ is the subset of gauge fixed configuration, and $\xi$ is the gauge fixing parameter. The Landau gauge $\xi=0$ is regarded as a specific limit of the Curci-Ferrari gauge.  In recent studies, the space-time average of the vacuum expectation value
\[
\frac{1}{V}\int_V \langle\frac{1}{2} tr {\mathcal A}_\mu(x){\mathcal A}^\mu(x)\rangle d^4 x
\]
is claimed to have gauge invariant meaning\cite{kondo, slav}.
In the Landau gauge QCD, the Faddeev-Popov(FP) gauge-fixing action is
\begin{equation}
S_{FP}=B^a\partial_\mu A^a_\mu+i\bar {\mathcal C}^a{\partial_\mu D_\mu}^{ab}{\mathcal C}^b,
\end{equation}
where the last term
$\bar {\mathcal C}^a{\partial_\mu D_\mu}^{ab}{\mathcal C}^b$, where $D_\mu^{ab}=\delta^{ab}\partial_\mu +gf^{acb}A^c_\mu$.

In analytical studies in the Curci-Ferrari gauge, presence of the ghost condensate $\langle f^{abc}c^b \bar c^c\rangle$ was discussed as the Overhauser effect in contrast to the $\langle f^{abc} c^b c^c\rangle$ or $\langle f^{abc}\bar c^b \bar c^c\rangle$ which are regarded as the BCS effect\cite{dvlsspvg}.

Since the Landau gauge is a specific limit of the Curci-Ferrari gauge, it is of interest to study the ghost propagator. In \cite{FN04}, we observed that in the SU(2) $\beta=2.1$ $16^4$ lattice, the expectation value of color off-diagonal ghost propagator $\langle \epsilon^{abc}\bar c^b c^c\rangle$ is consistent with 0 but the standard deviation of the color-diagonal ghost propagator has the momentum dependence of $\sigma(G^{aa}(q))\propto q^{-4}$. The investigation was extended by \cite{CMM} and this fluctuation was confirmed and although the expectation value of $\phi^a(q)=\epsilon^{abc}c^b \bar c^c$ is consistent with 0, the expectation value of its absolute value $|\phi^a(q)|$ was shown to behave as $q^{-4}$ and not zero. We extend this approach to unquenched MILC configurations.

In \cite{CMM}, the ghost condensate parameter $v$ and the Binder cumulant\cite{bind} of the color anti-symmetric ghost propagator was measured. In the Binder cumulant of an order parameter, renormalization factors cancel and one can extract the fixed point in the continuum limit by a suitable extrapolation.

In the Zwanziger's Lagrangian\cite{Zwa}, the color anti-symmetric ghost field $\phi^{bc}_\mu(x)$ leads to the mass gap equation
\begin{equation}
f^{abc}\langle A^{a\mu}(x)\phi^{bc}_\mu(x)\rangle=\frac{4(N_c^2-1) \gamma^2}{\sqrt 2 g^2}
\end{equation}
where $\gamma^2$ is the mass dimension two Gribov mass parameter\cite{gra}.
It is not evident that the Zwanziger's Lagrangian expresses the effective theory of the lattice Landau gauge QCD, but analytical calculation of the ghost propagator in two loop\cite{gra}, and the local composite operator approach\cite{dvlsspvg,cdglssv} suggest hints for solving entanglements in the confinement problem. 

In this paper we study the ghost propagator of quenched SU(2) $\beta=2.2$ $16^4$ lattice gauge fixed to the Landau gauge via parallel tempering(PT) method\cite{NF} and investigate the Binder cumulant. We extend the study to unquenched SU(3), using the MILC$_c$ configurations\cite{milc}.

Organization of the paper is as follows. In sect. II, we show definitions of the color diagonal and color anti-symmetric 
 ghost propagator on the lattice.  In sect.III, fluctuation of the ghost propagator of quenched and unquenched configurations
 are compared. In sect.IV, we compare the parameter $v$ of the ghost condensates from the color anti-symmetric ghost propagators
of quenched SU(2) PT configurations and unquenched MILC$_c$ configurations. In sect.V, the Binder cumulant of the color anti-symmetric 
ghost propagator of the quenched SU(2) and unquenched SU(3) are compared.  Summary and discussion are given in sect. VI.

\section{The ghost propagator}
The ghost propagator ${D_G}^{ab}(q^2)$ and the ghost dressing function $G^{ab}(q^2)$ is defined by the Fourier transform(FT) of the  expectation value of the inverse Faddeev-Popov operator $\cal  M=-\partial_\mu D_\mu$
\begin{eqnarray}
FT[{D_G}^{ab}(x,y)]&=&FT\langle {\rm tr} ( \Lambda^{a \dagger} \{({\cal  M}[U])^{-1}\}_{xy}
\Lambda^b  )\rangle,\nonumber\\
&=&{D_G}^{ab}(q^2)=\frac{G^{ab}(q^2)}{q^2}
\end{eqnarray}
where antihermitian SU(3) generator $\Lambda^a$ is normalized as ${\rm tr} \Lambda^{a\dagger} \Lambda^b=\delta^{ab}$.

 We measure
\begin{equation}
{D_G}^{ab}(q^2)=\left\langle tr\langle \Lambda^a q|{\mathcal M}[U]^{-1}|\Lambda^b q\rangle \right\rangle
\end{equation}
using the source vector $\displaystyle |\Lambda^a q\rangle= \frac{1}{\sqrt V}\Lambda^a e^{i q\cdot x}$.
We select the momentum $q_\mu$  to be directed along the diagonal of the lattice momentum space.  

In the approach of calculating the fourier transform of ${\mathcal M}^{-1}S^a_0(x)$\cite{Orsay}, 
compensation of hypercubic artefacts was necessary, but in our method the artifact-free momenta are selected and the translation invariance is fully utilized to improve the statistics.

The Faddeev-Popov operator ${\mathcal M}[U]=-\partial D[U]$ is defined with use of the covariant derivative as
\begin{equation}
D_\mu(U_{x,\mu})\phi=S(U_{x,\mu})\partial_\mu\phi+[A_{x,\mu},\bar\phi]
\end{equation}
where $\partial_\mu \phi=\phi(x+\mu)-\phi(x)$, $\bar\phi=\frac{1}{2}(\phi(x+\mu)+\phi(x))$. 
In the $U$-linear version, ($A_{x,\mu}=\frac{1}{2}(U_{x,\mu}-U_{x,\mu}^\dagger)|_{trlp}$ where $|_{trlp}$ means the traceless part) $S(U_{x,\mu})B_{x,\mu}$ is defined as
\begin{equation}
S(U_{x,\mu})B_{x,\mu}=\frac{1}{2}\{\frac{U_{x,\mu}+U_{x,\mu}^\dagger}{2},B_{x,\mu}\}|_{trlp}
\end{equation}
and in the $\log-U$ version, ($U_{x,\mu}=e^{A_{x,\mu}}$) 
\begin{equation}
S(U_{x,\mu})B_{x,\mu}=\frac{A_{x,\mu}}{2\tanh(A_{x,\mu}/2)}B_{x,\mu},
\end{equation}
where $A_{x,\mu}=adj A_{x,\mu}$\cite{NF}. 

In \cite{CMM}, the Faddeev-Popov operator was parametrized as
\begin{equation}
{\mathcal M}^{bc}_U(x,y)=\delta^{bc}{\mathcal S}(x,y)-f^{bcd}{\mathcal A}^d(x,y)
\end{equation}
The authors decomposed the inverse matrix ${D_G}^{bc}(x,y)=({\mathcal M}^{-1})^{bc}(x,y)$ into ${D_G}^{bc}_e(x,y)$ and ${D_G}^{bc}_o(x,y)$ i.e. the component containing even number of ${\mathcal A}$  odd number of ${\mathcal A}$, respectively. They derived the ghost propagator from $\displaystyle\frac{\delta^{bc}}{N_c^2-1}{D_G}^{bc}_e(x,y)$, and 
the color antisymmetric ghost propagator by multiplying ${\mathcal S}^{-1}{\mathcal A}$ to the color antisymmetric ghost propagator ${D_G}^{bc}_e(x,y)$ which contains perturbation series of even numbers of ${\mathcal A}$.  

 We do not adopt this procedure, but derive directly the color anti-symmetric ghost propagators by the conjugate gradient method.  The convergence condition on the series is set to less than a few \% in the ${\it l}_2$ norm. 

We define ${\mathcal M}=-\partial_\mu D_\mu$ and solve
\begin{equation}
-\partial_\mu D_\mu f_s^b({\bf x})=\frac{1}{\sqrt V}\Lambda^b \sin{\bf q}\cdot{\bf x}
\end{equation}
and
\begin{equation}
-\partial_\mu D_\mu f_c^b({\bf x})=\frac{1}{\sqrt V}\Lambda^b \cos{\bf q}\cdot{\bf x}.
\end{equation}
Then we calculate the overlap to get the color diagonal ghost propagator
\begin{eqnarray}
&&D_G(q)=\frac{1}{N_c^2-1}\frac{1}{V}\nonumber\\
&&\times\delta^{ab}(\langle \Lambda^a\cos{\bf q}\cdot{\bf x}|f_c^b({\bf x})\rangle+\langle \Lambda^a\sin{\bf q}\cdot{\bf x}|f_s^b({\bf x})\rangle)\nonumber\\
\end{eqnarray}
and color anti-symmetric ghost propagator
\begin{eqnarray}
&&\phi^c(q)=\frac{1}{\mathcal N}\frac{1}{V}\nonumber\\
&&\times f^{abc}(\langle \Lambda^a\cos{\bf q}\cdot{\bf x}|f_s^b({\bf x})\rangle-\langle \Lambda^a\sin{\bf q}\cdot{\bf x}|f_c^b({\bf x})\rangle)\nonumber\\
\end{eqnarray}
where ${\mathcal N}=2$ for SU(2) and 6 for SU(3).

\section{Fluctuation of the ghost propagator}
We present in the following subsections square and the absolute value of the color antisymmetric ghost propagators of the quenched SU(2) $\beta=2.2$ $16^4$ lattice, and compare the corresponding values of SU(2) Landau gauge QCD of larger samples\cite{CMM}. We measure also ghost propagators of quenched SU(3) $\beta=6.45$ $56^4$ lattice  and those of unquenched MILC$_c$ with lattice size $20^3\times 64$ and MILC$_f$ with lattice size $28^3\times 96$. We present square and the absolute value of color antisymmetric ghost propagator of MILC$_c$. 

\subsection{Quenched SU(2)}

We select momenta $q$ following the cylinder cut, and in the case of unquenched SU(3) $20^3\times 64$ lattice calculation, it takes about 260 iterations in the $q=0.2$GeV region but several iterations in the $q=4$GeV region.
The average of color anti-symmetric ghost propagator $\phi^c(q)$ is consistent with 0 but the average of its square $\phi^c(q)^2$ has a non-vanishing value.
We define 
\begin{equation}
\vec \phi(q)^2=\frac{1}{N_c^2-1}\sum_{c}\phi^c(q)^2
\end{equation}
\begin{figure}[htb]
\includegraphics[width=7.2cm,angle=0,clip]{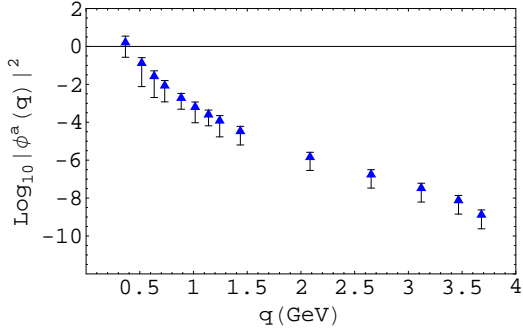}
\caption{Log of color anti-symmetric ghost propagator squared $\log_{10}[\phi(q)^2]$ as the function of $q$(GeV). $\beta=2.2$, $16^4$ PT gauge fixing. (Color online)}\label{phi_ptn}
\end{figure}

\begin{figure}[htb]
\includegraphics[width=7.2cm,angle=0,clip]{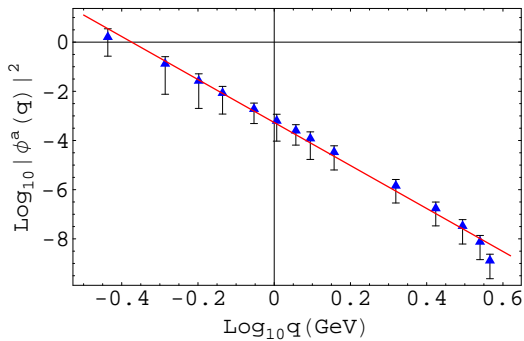}
\caption{Log of the color anti-symmetric ghost propagator squared $\log_{10}[\phi(q)^2]$ as the function of $\log_{10}[q$(GeV)]. $\beta=2.2$, $16^4$ PT gauge fixing. (Color online)}\label{phi_pt}
\end{figure}

\begin{figure}[htb]
\includegraphics[width=7.2cm,angle=0,clip]{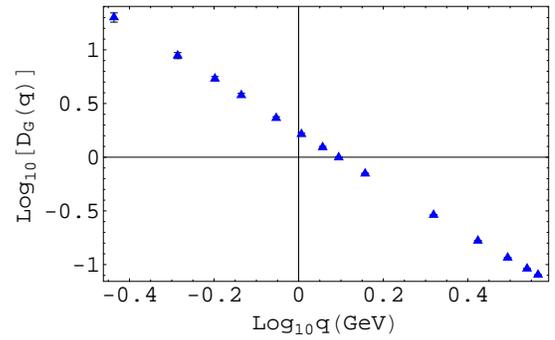}
\caption{Log of the ghost propagator $\log_{10}[D_G(q)]$ as the function of $\log_{10}[q$(GeV)]. $\beta=2.2$, $16^4$ PT gauge fixing. (Color online)}\label{gh_pt}
\end{figure}

The log of $\vec \phi(q)^2$ of $\beta=2.2$ $16^4$ lattice gauge 
fixed by the PT method (67 samples)
is shown in Fig.\ref{phi_ptn}. The corresponding log-log plots (Fig.\ref{phi_pt}) is to be compared with  that of the ghost 
propagator $D_G(q)$ (Fig.\ref{gh_pt}).


The infrared singularity of the standard deviation of the color anti-symmetric ghost dressing function and the color diagonal ghost dressing function are $q^{-4.4}$ and $q^{-4.5}$, respectively.

\subsection{Quenched SU(3)}
\begin{figure}[htb]
\includegraphics[width=7.2cm,angle=0,clip]{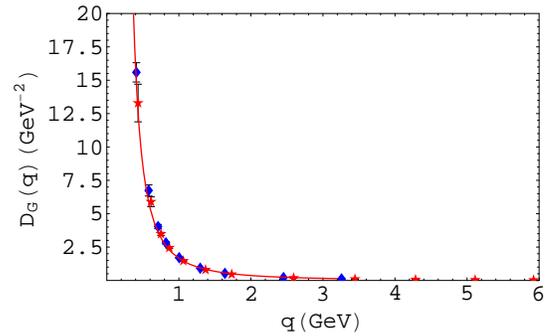}
\caption{The ghost propagator as the function of the momentum $q$(GeV). $\beta=6.45$, $56^4$(stars) and $\beta=6.4, 56^4$(filled diamonds) in the $\log U$ definition. Solid line is the pQCD fit in $\widetilde{MOM}$ scheme\cite{FN05}. (Color online)}\label{gh645_64}
\end{figure}

\begin{figure}[htb]
\includegraphics[width=7.2cm,angle=0,clip]{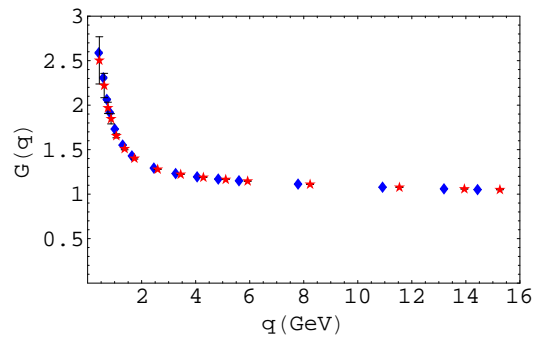}
\caption{The ghost dressing function as the function of the momentum $q$(GeV). $\beta=6.45$, $56^4$(stars) and $\beta=6.4, 56^4$(filled diamonds) in the $\log-U$ definition. (Color online)}\label{ghd645_64}
\end{figure}

In FIG.\ref{gh645_64} we show the color diagonal ghost propagator of quenched SU(3) with $\beta=6.4$ ($1/a=3.66$GeV) and $\beta=6.45$ ($1/a=3.8697$GeV) on $56^4$ lattice. The corresponding ghost dressing function are in FIG.\ref{ghd645_64}.

The standard deviation of the color-diagonal ghost propagator of $\beta=6.45$
multiplied by $(qa)^4$ is almost constant in $q>1$GeV region but in the $q<0.5$GeV region it is enhanced as compared to the value at $q>1$GeV region.

The $\log-\log$ plot of the standard deviation of the color-diagonal ghost dressing  function of $\beta=6.45$ in the $q<1$GeV region behaves as
\begin{equation}
\sigma(G(q))\propto q^{-2.8(1)}.
\end{equation}

\subsection{Unquenched SU(3)}
\begin{figure}[htb]
\begin{center}
\includegraphics[width=7.2cm,angle=0,clip]{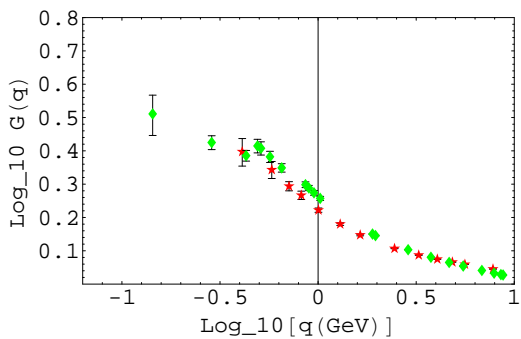}
\end{center}
\caption{Log of the ghost dressing function $\log_{10} G(q)$ as a function of  $\log_{10} q(\rm{GeV})$ of MILC$_f$ $\beta_{imp}={7.09}$ (diamonds) and that of quenched $\beta=6.45$ $56^4$ (stars). (Color online)}\label{ghdmilc_645}
\end{figure}

In \cite{FN05} we showed lattice results of color diagonal ghost dressing function of unquenched JLQCD/CP-PACS and MILC. In these simulations the length of the time axis is longer than the spacial axes and the ghost propagator of low momentum region is extended. In FIG.\ref{ghdmilc_645} the log-log plot of the ghost dressing function of the MILC$_f$ $\beta_{imp}=7.09$ on $28^3\times 96$ lattice and that of quenched $\beta=6.45$ on $56^4$ lattice are shown. We observe suppression of the ghost propagator in the infrared region in the asymmetric lattice\cite{unquench,lat05_a}. Systematic deviation of ghost propagator and gluon propagator of asymmetric lattice from those of symmetric lattice is recently confirmed in the large 3-dimensional SU(2) lattice\cite{CM}. The suppression in the infrared of the unquenched data may not be due to the presence of quarks but due to the geometry of the lattice.

\begin{figure}[htb]
\begin{center}
\includegraphics[width=7.2cm,angle=0,clip]{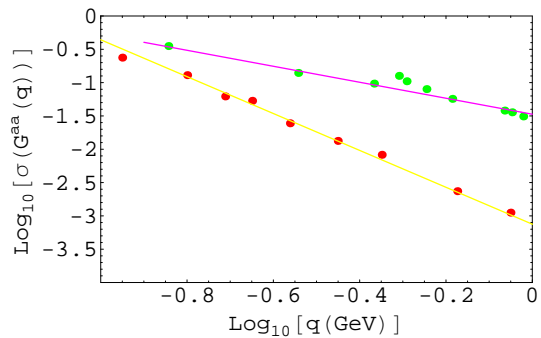}
\end{center}
\caption{Log of the standard deviation $\log_{10} \sigma(G(q))$ as a function of  $\log_{10} q(\rm{GeV})$ of MILC$_f$ $\beta_{imp}={7.09}$ (upper points) and that of quenched $\beta=6.45$ $56^4$ (lower points). (Color online)}\label{sigma_g}
\end{figure}

There are difference in the momentum dependence of the standard deviation of the color diagonal ghost dressing function of unquenched MILC$_f$, $\beta_{imp}=7.09$ on $28^3\times 96$ lattice and the quenched $\beta=6.45$ on $56^4$ lattice in the infrared region as shown in FIG.\ref{sigma_g}.
Since the sample size is different, the absolute value of the standard deviation is not meaningful, but the strength of the fluctuation defind by the slope influences the infrared behavior of the running coupling etc.

The momentum dependence of the standard deviation of the color diagonal ghost dressing function of MILC$_f$ is dramatically less singular than that of quenched configuration. We observed
\begin{equation}
\sigma(G(q))\propto q^{-1.1(1)}.
\end{equation}

\begin{figure}[htb]
\includegraphics[width=7.2cm,angle=0,clip]{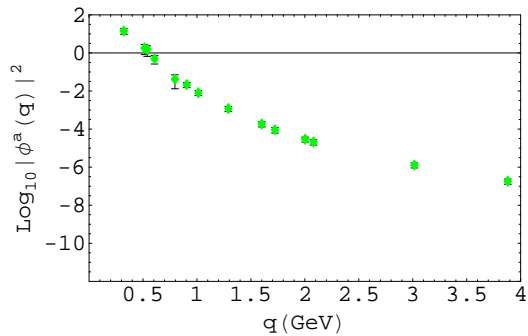}
\caption{Log of the color anti-symmetric ghost propagator squared $\log_{10}[\phi(q)^2]$ as the function of $q$(GeV). $\beta_{imp}=6.83$ and 6.76, $20^3\times 64$ MILC$_c$. (Color online)}\label{phi_milc}
\end{figure}

The color anti-symmetric ghost propagator of MILC$_c$ (21 samples) is shown in Fig.\ref{phi_milc}. By comparing with FIG.\ref{phi_ptn}, we observe decrease of the slope. 

\section{The ghost condensate}
In the case of SU(2), the ghost condensate appears in the color anti-symmetric ghost propagator ${D_G}^{bc}_o(q)$ related to $\phi^a(q)$ through
\begin{equation}
\phi^a(q)=-i\frac{f^{abc}}{2}{D_G}^{bc}(q^2)=-i\frac{f^{abc}}{2}{D_G}^{bc}_o(q)
\end{equation}
and
\begin{equation}
{D_G}^{bc}_o(q)=i\frac{r/L^2+v}{q^4+v^2}\epsilon^{bc},
\end{equation}
where $\epsilon^{bc}$ is an anti-symmetric tensor, i.e. when $a=3$, $b$ and $c=1, 2$. In general, we parametrize the average of $|\phi^a(q)|$ as
\begin{equation}
\frac{1}{N_c^2-1}\sum_a|\phi^a(q)|= \frac{r/L^2+v}{q^4+v^2}
\end{equation}

Here $L$ is the lattice size and the parameter $r/L^2$ is the correction from the finite size effect.

\subsection{Quenched SU(2)}
In \cite{CMM}, the fitting parameter $r$ of $|\vec\phi^a(q)|$ (color antisymmetric ghost propagator) on the lattice was derived from
\begin{equation}
\frac{1}{3}\sum_a\frac{L^2}{\cos(\pi \bar q/L)}|\phi^a(q)|=\frac{r}{q^z},
\end{equation}
in which $\bar q=0,1,\cdots L$. 

\begin{figure}[htb]
\includegraphics[width=7.2cm,angle=0,clip]{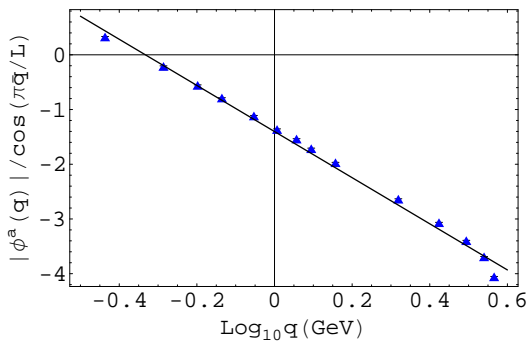}
\caption{Log of $|\phi^a(q)|$ (color anti-symmetric ghost propagator) devided by $\cos(\pi\bar q a/L)$ as the function of  $\log_{10}(q$(GeV)) of SU(2) PT samples. (Color online)}\label{su2gha_1}
\end{figure}

Our fit of $\displaystyle \frac{|\phi^a(q)|}{\cos(\pi\bar q a/L)}$ of PT samples using $r=10.13, z=4.215$ is shown in FIG.\ref{su2gha_1}.

\begin{figure}[htb]
\includegraphics[width=7.2cm,angle=0,clip]{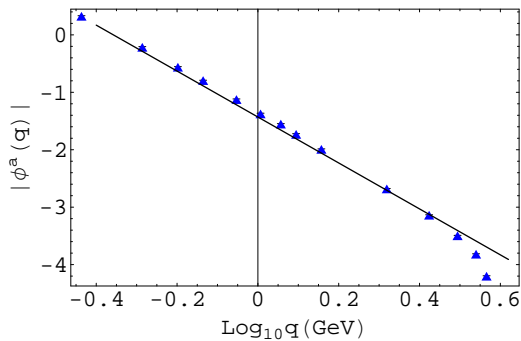}
\caption{Log of the absolute value $|\phi^a(q)|$ (color anti-symmetric ghost propagator) as the function of  $\log_{10}(q$(GeV)) of SU(2) PT samples. (Color online)}\label{su2gha_2}
\end{figure}

Using $r=10.13, L=16$, the fitting parameter $v$ of $|\phi(q)|$ is found to be -0.002GeV$^2$ and is consistent with 0. The fit is shown in FIG.\ref{su2gha_2}.

\subsection{Unquenched SU(3)}
\begin{figure}[htb]
\includegraphics[width=7.2cm,angle=0,clip]{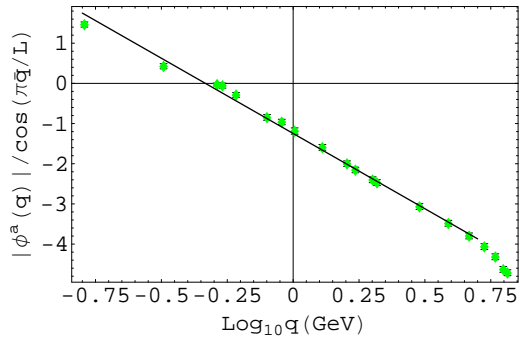}
\caption{Log of $|\vec\phi(q)|/\cos(\pi\bar q a/L)$ (color anti-symmetric ghost propagator) as the function of  $\log_{10}(q$(GeV)) of MILC$_c$ samples. (Color online) }\label{milcgha_1}
\end{figure}

As in the SU(2) PT samples, we performed the fit of the parameter $v$ for the MILC$_c$ samples. We first fit the $\log$ of $|\phi^a(q)|$(color anti-symmetric ghost propagator) devided by $\cos(\pi\bar q a/L)$ using $L=\sqrt {20^3\times 64}$ and obtained $r=40.5, z=3.75$, as shown in FIG.\ref{milcgha_1}. The parameter $r$ and $z$ for the fit of $\vec \phi^2(q)/\cos^2(\pi \bar q/L)$ are $r=36.5, z=7.5$.

\begin{figure}[htb]
\includegraphics[width=7.2cm,angle=0,clip]{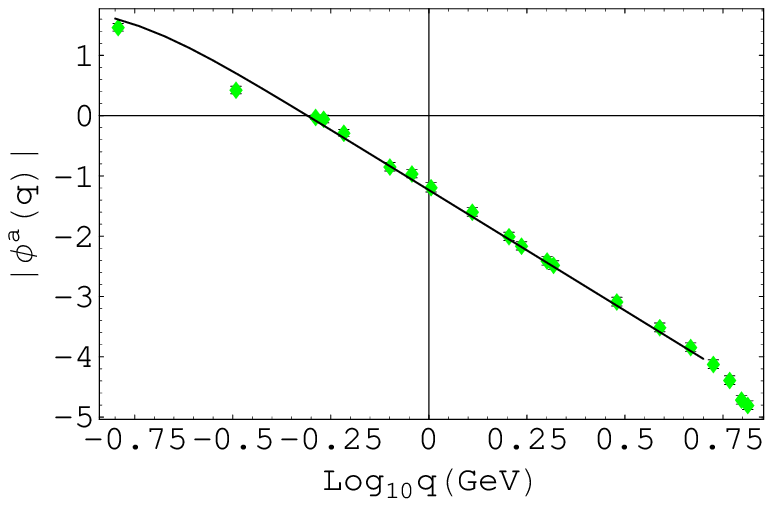}
\caption{Log of the absolute value $|\vec\phi(q)|$ (color anti-symmetric ghost propagator) as the function of  $\log_{10}(q$(GeV)) of MILC$_c$ samples. (Color online)}\label{milcgha_2}
\end{figure}

Our fit of $|\phi(q)|$ ignoring two lowest momentum points and using $r=40.51$ gives $v=0.0020$GeV$^2$, which is small but positive. When the two lowest momentum points are included, $v$ decreases to -0.0005 but $\chi^2/d.o.f.$ increases.
The former fit is shown in FIG.\ref{milcgha_2}.

We fitted also $\log_{10}\vec \phi^2(q)$, where 
\begin{equation}
\vec \phi^2(q)=\frac{1}{N_c^2-1}\sum_a \phi^a(q)^2= (\frac{r/L^2+v}{q^4+v^2})^2
\end{equation}

\begin{figure}[htb]
\includegraphics[width=7.2cm,angle=0,clip]{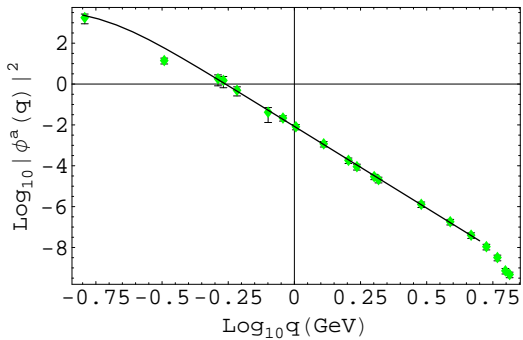}
\caption{Log of the $\vec\phi(q)^2$ (color anti-symmetric ghost propagator squared) as the function of  $\log_{10}(q$(GeV)) of MILC$_c$ samples. (Color online)}\label{milcgha_3}
\end{figure}

The fit with $r=40.5$, $v=0.035$GeV$^2$ is shown in FIG.\ref{milcgha_3}. The fit with $r=36.5, v=0.041$GeV$^2$ is not distinguished from this figure.

\section{Binder cumulant}
Two decades ago Binder\cite{bind} showed cumulants of the order parameter yields non-trivial fixed-point values. The theory was applied to the Ising model in which the magnetization $M$ is the order parameter\cite{pr, bptr} and the cumulant was defined as
\begin{equation}\label{bind_ising}
B=\frac{1}{2}(3-\frac{\langle M^4\rangle}{\langle M^2\rangle^2}).
\end{equation}
When the distribution of $M$ is given by the 1 dimensional gaussian distribution, one finds
\begin{equation}
\frac{\langle M^4\rangle}{\langle M^2\rangle^2}=3.
\end{equation}
and $B$ becomes 0.

In SU(2) and SU(3) lattice QCD, deconfinement phase transition was studied by measuring
\begin{equation}
g=\frac{\langle P^4\rangle}{\langle P^2\rangle^2}-3
\end{equation}
using the Polyakov line data $P$ as the order parameter\cite{abs,karsch}.

Since the color anti-symmetric ghost propagator could be an order parameter of the system, the authors of \cite{CMM} considered its Binder cumulant defined as
\begin{equation}
U(q)=1-\frac{\langle\vec\phi(q)^4\rangle}{3\langle\vec\phi(q)^2\rangle^2}.\label{bind}
\end{equation}

We measure 
\begin{eqnarray}
&&\vec\phi(q)^2=\frac{1}{N_c^2-1}\sum_a [\frac{1}{\mathcal N}\nonumber\\
&&\times \frac{f^{abc}}{V}\left(\langle\Lambda^{b}\cos{\bf q}\cdot{\bf x}|{\mathcal M}^{-1}| \Lambda^{c}\sin{\bf q}\cdot{\bf x}\rangle\right.\nonumber\\
&&\left.-\langle\Lambda^{b}\sin{\bf q}\cdot{\bf x}|{\mathcal M}^{-1}| \Lambda^{c}\cos{\bf q}\cdot{\bf x}\rangle \right)]^2
\end{eqnarray}
and
\begin{eqnarray}
\vec\phi(q)^4&=&(\frac{1}{N_c^2-1}\sum_a [\frac{1}{\mathcal N}\nonumber\\
&&\times \frac{f^{abc}}{V}\left(\langle\Lambda^{b}\cos{\bf q}\cdot{\bf x}|{\mathcal M}^{-1}| \Lambda^{c}\sin{\bf q}\cdot{\bf x}\rangle\right.\nonumber\\
&&\left.-\langle\Lambda^{b}\sin{\bf q}\cdot{\bf x}|{\mathcal M}^{-1}| \Lambda^{c}\cos{\bf q}\cdot{\bf x}\rangle\right)]^2)^2.
\end{eqnarray}

In arbitrary d-dimensional space, corresponding expectation value for d-dimensional gaussian distribution becomes
\begin{equation}
\frac{\langle \vec\phi^4\rangle}{\langle \vec\phi^2\rangle^2}=\frac{d+2}{d}.
\end{equation}
Thus a natural extension to d-dimensional vector variable is
\begin{equation}
\tilde U(q)=\frac{\langle \vec\phi^4\rangle}{\langle \vec\phi^2\rangle^2}-\frac{d+2}{d}
\end{equation} 
which becomes 0 in the system with gaussian distribution whose symmetry is not broken.

When the symmetry of the system is broken, as in the Ising model at the 0 temperature, the ratio of $\langle\vec\phi(q)^2\rangle^2$ and $\langle\vec\phi(q)^4\rangle$  becomes 1 and $\tilde U(q)$ becomes $-\displaystyle\frac{2}{d}$. It corresponds to the 0 temperature fixed point.

\subsection{Quenched SU(2)}

\begin{figure}[htb]
\includegraphics[width=7.2cm,angle=0,clip]{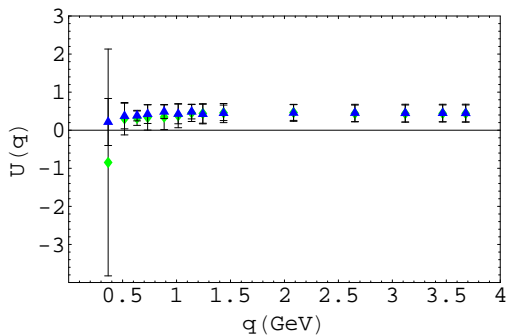}
\caption{The momentum dependence of Binder cumulant $U(q)$ of SU(2), $\beta=2.2$, $a=1.07$GeV$^{-1}$ of PT samples and first copy samples.
(color online) }\label{u_pt_ptr}
\end{figure}

We measure the Binder cumulant $U(q)$ of the quenched SU(2) $16^4$ $\beta=2.2$, $a=1.07$GeV$^{-1}$  configurations (67 samples) produced by the PT Landau gauge fixing and the corresponding first copy\cite{NF,FN04}. 
 The $q$ dependence of the $U(q)$ of PT gauge fixed samples and the first copies are shown in FIG. \ref{u_pt_ptr}. The infrared fluctuation is large in the first copy but it is reduced in the PT gauge fixed samples. It implies a Gribov copy effect in the infrared region\cite{muell}. 
The average over $q>0.5$GeV becomes $U(q)=0.45(2)$. This value is comparable to that of \cite{CMM} obtained by 10000 samples using symmetric momentum $q_1=q_2=q_3=q_4\ne 0$. In \cite{CMM}, the value $U(q)$ between 0 and 2/3 was interpreted as a system deviating from the gaussian distribution. However, since $\vec\phi(q)$ is a 3-dimensional vector, it would not be appropriate to treat it as an 1 dimensional object. The value 0.45 is very close to 
\begin{equation}
U(q)\sim 1-\frac{d+2}{3d}=\frac{4}{9}, 
\end{equation}
or 
\begin{equation}\label{u_tilde}
 \tilde U(q)\sim \frac{5}{3}-\frac{d+2}{d}= 0
\end{equation}
corresponding to the 3-d gaussian distribution.

\subsection{Unquenched SU(3)}

\begin{figure}[htb]
\includegraphics[width=7.2cm,angle=0,clip]{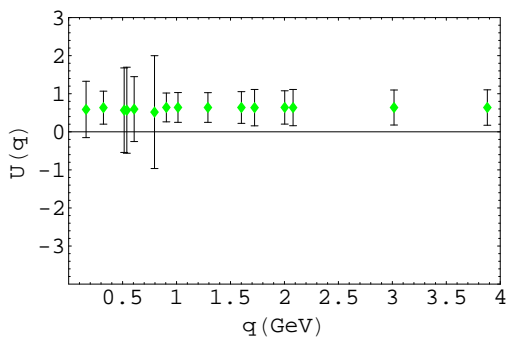}
\caption{The momentum dependence of Binder cumulant $U(q)$ of unquenched SU(3),  $\beta_{imp}=6.83$ and 6.76, $a=1.64$GeV$^{-1}$ MILC$_c$.
(color online) }\label{u_milc}
\end{figure}

We measured the Binder cumulant of the color anti-symmetric ghost propagator of MILC$_c$.  We observed qualitatively different features from quenched SU(2). 
An average of 9 $\beta_{imp}=6.76$ samples and 12 $\beta_{imp}=6.83$ samples of MILC$_c$ is shown in Fig.\ref{u_milc}.  When the $\vec \phi$ is
distributed as a Gaussian vector in eight dimensional space, 
$U(q)=1-{10\over{3\cdot 8}}=0.58$. Data of Fig.\ref{u_milc}
suggests that $U(q)$ is slightly larger than the 0.58, and that the shift from Gaussian distribution of $\beta_{imp}=6.76$ samples with light bare quark mass $m_0=11.5MeV$
is larger than that of $\beta_{imp}= 6.83$ samples with heavier bare quark mass $m_0=65.7MeV$.

A qualitative difference of unquenched SU(3) (Fig.\ref{u_milc}) from quenched SU(2) (Fig.\ref{u_pt_ptr}) is the smallness of the fluctuation 
at the lowest and next to the lowest momentum point [$q=(0,0,0,1)$ and (0,0,0,2)]. Relatively large fluctuation exists when one of the spacial components of $q$ is 1 and other components are 0. The difference from the quenched SU(2) could be due to the improvement in the Asqtad action used in the MILC$_c$ and/or the presence of dynamical fermions.

\section{Summary and Discussion}
We presented color diagonal ghost propagator of quenched $\beta=6.45$ $56^4$ lattice and those of unquenched MILC$_c$ $20^3\times 64$ and MILC$_f$ $28^3\times 96$ configurations.  The momentum dependence of standard deviation of the color diagonal ghost dressing function of the unquenched configurations is less singular than that of the quenched configurations. 
The standard deviation and the mean value of statistical distribution is important for determining the nature of the ensemble.  

The ghost pair creation operator in the BCS channel is expected to behave as the order parameter, and in the Landau gauge, where ghost pair creation is absent, the ghost anti-ghost pair creation in the Overhauser channel was speculated to become an order parameter. The parameter $v$ of LCO approach that characterize the ghost condensate was compatible with 0 in the SU(2) PT samples. In the unquenched SU(3) MILC$_c$  samples, we found small but positive value of $v$. Uncertainty on $v$ comes mainly from that of $r$, where finite size effect is crucial. We need to extend the analysis to larger lattices to get the definite conclusion. 

We showed that the Binder cumulant which measures the fluctuation of the ghost propagator differ between quenched and unquenched configurations.
We confirmed that the Binder cumulant $U(q)$ of the color anti-symmetric ghost propagator of SU(2) obtained by 10000 samples\cite{CMM} $U(q)\sim \frac{4}{9}$ is consistent with that obtained by using PT gauge fixed samples.   In 3-dimensional system,  this data can be interpreted as $\tilde U(q)$ defined as eq.(\ref{u_tilde})$\sim 0$, i.e. the color symmetry is not broken and that the system is in the random phase.  When the system is ordered, a certain direction in the color space will be selected and the Binder cumulant would deviate from the value expected by the gaussian distribution. The data of quenched SU(2) $\beta=2.2$ do not show this tendency but the unquenched SU(3) show deviation from the gaussian distribution.  Whether it implies the precourser of the ghost condensation in the unquenched QCD is not evident. It would be interesting to extend the analysis to finite temperature and study qualitative differences.

The ghost condensates and the $A^2$ condensates are expected to be related by the on-shell BRST symmetry. The observables of lattice Landau gauge in 1-3GeV region suggests presence of $A^2$ condensates. The larger standard deviation of the SU(3) quenched ghost propagator as compared to the unquenched ghost propagator may imply that the ghost propagator is more random in the quenched samples. 
The fluctuation of the ghost propagator could be the main cause of suppression of the running coupling in the infrared and saturation of the Kugo-Ojima parameter $c$ at about 0.8 in the quenched approximation.
It is likely that the fermion field reduces the fluctuation of the color diagonal ghost propagator in the infrared, and renders the Kugo-Ojima parameter $c$ close to 1.
 
We think infrared suppression of the running coupling of unquenched SU(3) measured by eq.(\ref{alpha}) presented in \cite{unquench} is a finite size effect. In the process of measuring the ghost propagator for the running coupling, 
 we observed large fluctuations of the norm and random orientation of the vector in adjoint color space i.e. weakening of the color diagonal structure of the ghost propagator in the infrared. Concerning the fixed points of the running coupling, Wilson\cite{Wilson} noted in 1971 that the renormalization group flow of the coupling could approach limit cycles which are more elaborate than simple isolated fixed points. A possibility of complicated fractal structure in fixed points  was discussed in \cite{MN}. 
The Zamolodchikov's c-theorem in two dimensional conformal field theory\cite{Zam}, however, excludes the limit cycle structure of the infrared fixed points. In four dimensional QCD, the situation is obscure\cite{Card}. To clarify the nature of the infrared fixed points, it is necessary to investigate the continuum limit of the lattice Landau gauge QCD via systematic studies of finite size effects and the Gribov copy effects.

\begin{acknowledgments}
 This work was supported by the KEK supercomputing project 04-106 and 05-128. The numerical calculation was performed also at the Cyber Media Center of Osaka Universitry. H.N. is supported by the MEXT grant in aid of scientific research in priority area No.13135210.
\end{acknowledgments}

\end{document}